# Interactive computer simulations of electrokinetic's physics phenomena

Gabriel Murariu[1]

[1]Faculty of Sciences,
*Dunarea de Jos* University, Galați, Romania
`Gabriel.Murariu@ugal.ro`

*Abstract* — In our days, the necessity of laboratory apparatus accustoming by building up specific software objects for studying the virtual evolution of physical phenomena is a major request. In this respect, the aim of the present paper is to present a developed set of electricity interactive simulation computer applications. By following an Object Oriented Programming technology, the necessary software objects are described. Finally, specific examples are extensively presented as well as some screen-shots of the applications interface.

## 1. Introduction

In the past few years, many physical phenomena are studied in a new way: using computer simulation programs. This strategy means a very powerful tool in understanding the concepts and relations between physical systems parts. The necessity of this way of physics study is needed by a couple of arguments. One of the most important is generated by the present trend in educational world: every student has a personal knowledge acquisition velocity and, this implies that every student has to get a "personal teacher". For such personalized educational process must be implied the computer. Using this approach the educational expected results for every student could be achieved.

In this way, can be remarked anterior papers [1-3], which depict this subject in different domains. The main problems of this strategy are revealed and very interesting results are presented. Many specific aspects are analyzed in an attractive mode. Noteworthy progresses are made in the computer assistance learning process [4-7]. Some characteristic are developed in an interactive way [8, 9] or using INTERNET technology [4, 6, 7].

On the other hand, considering the specific aspects of physics science, the permanent increasing complexity of measurement devices requests, in most of the cases, a preliminary training time for a briefly put up. Otherwise, the measurement process can be strongly perturbed and the results can contain errors. This important aspect can be solved very simple if the students could use training virtual instruments. In additions, many metrological aspects can be studied individually using this kind of devices. Subjects as the sensibility of apparatus or the measurement domains choosing can be discussed in a comparative mode: theoretical and practical. These aspects were few dis-cussed. The present paper tries to explain some specific results in this kind of approaching.

I do not generalize the using of the computer in the physics classes. The virtual experiments have not to substitute the real experiences. The function of the software applications is just a tool for teacher's help.

However, implying the computer in the school's activity can be offered a set of recompenses. An important advantage of this vision is given by the low-cost for using this kind of physics study. Now, can be simulated and studied a great series of physics phenomena without implying expensive resources [9-13]. On the other hand, everyone can get a personal virtual laboratory and a personal instruction teacher [9, 10, 12, 14].

However, this way implies a huge work volume to realize and build a set of computer applications. In present, teams of physicists and computer programmers raise this work. Many software companies are working in this direction.

For these computer programs it was used an old C++ compiler and the presented applications are from three different generations. It was followed an Object Oriented Programming technology and the developed structure of these software objects are further presented in an extensive way.

## 2. Applications' structure

In this paper it is presented a part of programs used in the first instruction level in electricity laboratory classes. To avoid a large exposure, the application's structure can briefly be described by simple enumerating the involved object. In the present set of electrical physics phenomena applications were used two software objects: electrical circuit network and analogical measurement device.

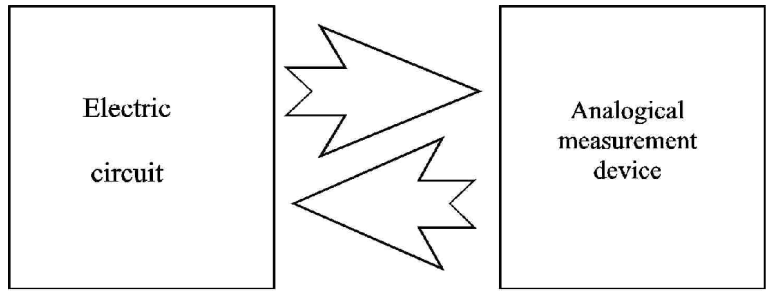

Figure 1 – Application's objects

The aim is to illustrate the interdependence between these two important components: electric circuit and measurement appliance. Students should be able to understand this mutual influence connecting two system parts. The considered structure is described in Figure 1.

The first object model describes the electrical network and was used a typical graph model. It will not discuss more detailed this application's part. Now are a lot of very powerful commercial software applications, which are designed to realize this kind of analysis. Thus, it will be presented only the particular part of designing work, spotlighting further only the specific aspects.

The second built object describes the measurement device. In order to succeed in training activity, it was chosen to consider analogical devices. This option was required by the observed difficulties in using this kind of laboratory device units.

*INTERACTIVE COMPUTER SIMULATIONS OF ELECTRICAL PHYSICS PHENOMENA*

This object realization requested a sequence of software structured objects. The generic list is presented in Figure 2 and the complete software procedure listing needs about fifty pages. In a few words it will be presented some specific aspects about the structure of this application's part. The presentation is made in a general view in order to be used in other physics software applications.

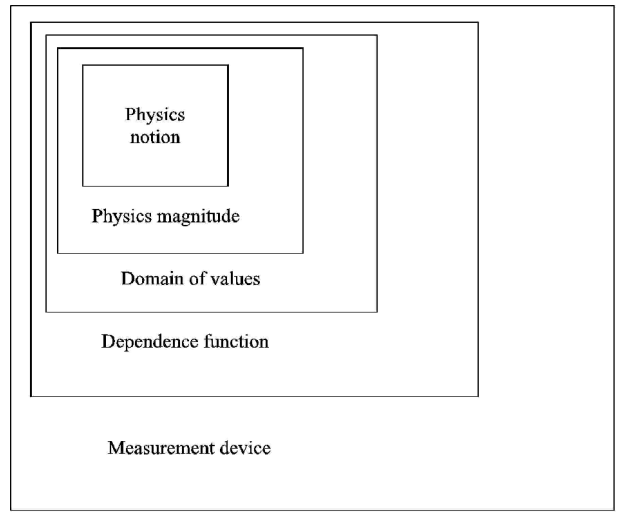

Figure 2 - The generic list of software objects

The first included object was the physics magnitude - a structure class with a character string field for the name, a double field for value and, of course, some integer fields for the state of the object.

    *class Magnitude*
    *{ double Value;// field for magnitude*
    *int State[];// integer fields for memorizing the actual state*
    *char\* Name;// filed for name*
    *// Implicit constructor procedure*
    ***public:*** *Magnitude ()*
    *{....}*
    *// Constructor procedure with a real parameter*
    *Magnitude ( double ax)*
    *{......}*

Further, using the class *Magnitude*, was built a new class named *Group* in order to create a software magnitude set of linked physics values.

    *class Group*
    *{ Magnitude Value;// object for measured magnitude*
    *Magnitude Divisions;// object for memorizing the indication*
    ***public:***
    *Group(double ValueCrt, double DivisionsCrt)*
    *{Value.ModifValue(ValueCrt);*
    *Divisions.ModifValue(DivisionsCrt);....}*
    *// Implicit constructor procedure*
    *Group()*
    *{Value.ModifValue(0);*
    *Value.ModifState(0);*
    *Divisions.ModifValue(0);...}*

The next object is named *AparatG* and realizes a generic structure for a great type number of apparatus. In this way can be modeled linear devices, logarithmic, etc. Some gauge constants are always needed in order to calibrate the values' domains.

```
class AparatG
{ Group ValueMinima;// retains the minimum value
Group ValueMaxima; // retains the minimum value
Magnitude ConstantC1;// retains a needed gauge constant
Magnitude ConstantC2; // retains a needed gauge constant
int State[];
private:
   double FunctieTip_1_C1(double y1, double x1, double y2, double x2)
       {…….;}
public:
   // Constructors
   AparatG();
```

The next level object is named *Aparat* and is a primordial object built in order to realize the simplest model of an analogical measurement device. This class contains some fields for the minimum measured value, for the maximum measured value and for the number of scale divisions and for the values' domain order.

```
class Aparat
{ Magnitude ValueMinima; //
  Magnitude ValueMaxima;//
  Magnitude OrderOf Magnitude;
  int NrDivisionsOnTheScale; //
Magnitude ValueCurent;//
   ...
public:
// Constructors of the object Aparat
Aparat()
   {……}
```

These initial construction objects are used further to build more complex units. Further is presented one of them named *AparatCircular*. It contains all the elements needed for the graphical interface. The first eleven field elements are involved in the order to build the interface description parts and the last five are needed for the graphical representation.

```
class AparatCircular
{ Aparat A;
Object Fig;
char* LabelMagnitude; // is a label for the name of the measured magnitude
char* LabelDivisions; // is a label for the name of the divisions number
char* LabelUnitsMeasureDown; // is a label for graphical domain divisions' down limit
char* LabelUnitsMeasureUp; // is a label for graphical domain divisions' up limit
List ListScale; // is a list of scale domains
char* WindowValueCurent; // is a label for the current division value
char* WindowDivisionsValueCurent; // is a label for the current value
char* WindowDomainCurent;
// Representation Unit
double AngleValueMinima; // is a value for the minimum division angle
```

*INTERACTIVE COMPUTER SIMULATIONS OF ELECTRICAL PHYSICS PHENOMENA*```
        double AngleValueMaximal; // is a value for the maximum division angle
        double AngleValueCurent; // is a value for the actual division angle
        .......

        //Center parameters
        int XStart;
        int YStart;
        // Representation parameters
        int Thikness;

    public:
        AparatCircular()
        {........}
```

These structured objects were used to realize simulation software applications for the electricity physics laboratory.

Further it will be presented a couple of software applications used in the first electricity laboratory classes. These programs were made in order to offer a simple aid for students. For each one is presented the generic electric scheme in order to be more enlightening.

Some general aspects can be expressed: each application has a main menu, each appliance may present help dialogs and correspondent theory paragraphs if the students are asking assist. Using the main menu can be modified the specific parameters for each appliance. Utilizing the help modulus, students could discovery the use.

## 3. Examples

The first presented program describes a well-known method of resistance measuring using a voltmeter and a current measure device. This application can be used in order to accustom the students with the connectivity mode of the used measure devices. The electric appliance scheme is presented in Figure 3 and the interface is given in Figure 4.

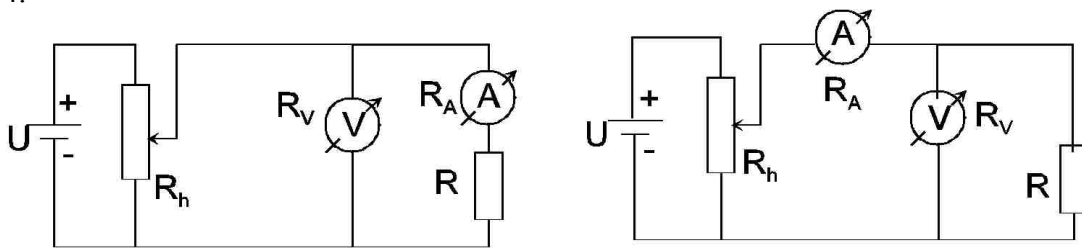

Figure 3 - The electric appliance for resistance measuring

This program was made in order to offer a first stage virtual experimental device to understand the influence of the connected apparatus in the electric circuit. Aspects like internal resistance influence on the apparatus' sensibility or measurement domain selection can be studied in an alternative mode.

The application is an advisor in the same time, offering recommendations during the "experiment", mainly in the domain surpassing times.



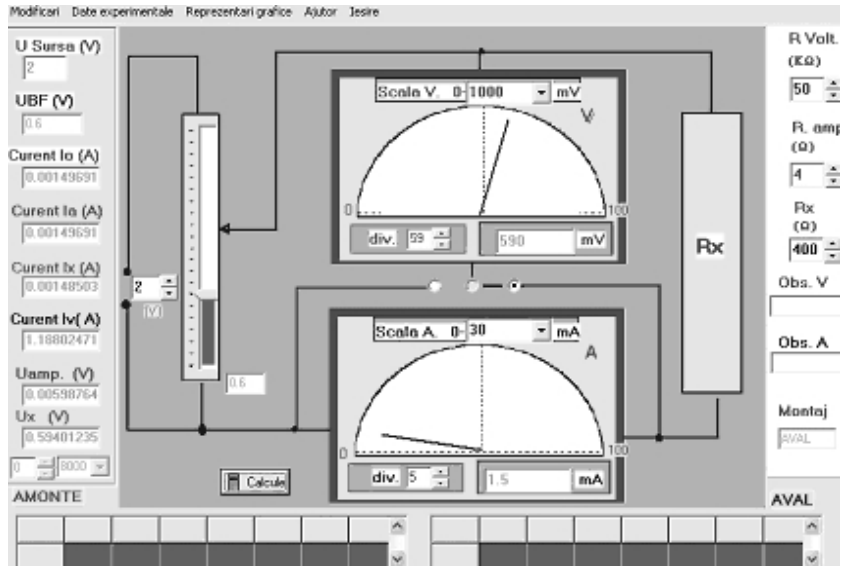

Figure 4 - The interface for resistance measuring application

The next example is one of the second-generation application software. It was designed to permit a Wheatstone assembly study. It was requested by the observation that students have to obtain some experimental skills before the effective experiment. This application is a good tool in order to succeed in obtaining such needed habits. The experimental data obtained without an anterior training time are more precise less. One aspect tracked during the experiment is to evaluate the sensibility of the measurement appliance.

The device sensibility is considered as

$$S = \frac{\Delta \theta}{\Delta R_x} \tag{1}$$

where $\Delta\theta$ is the galvanometer's indication change and $\Delta R_x$ is the correspondent measured resistance modification. This magnitude is not constant on the entire domain of values. Students should estimate the sensibility value for different cases in order to conclude on the best choice of circuit configuration.

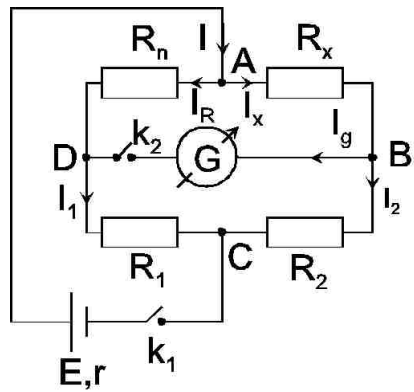

Figure 5 - The electric appliance for Wheatstone assembly



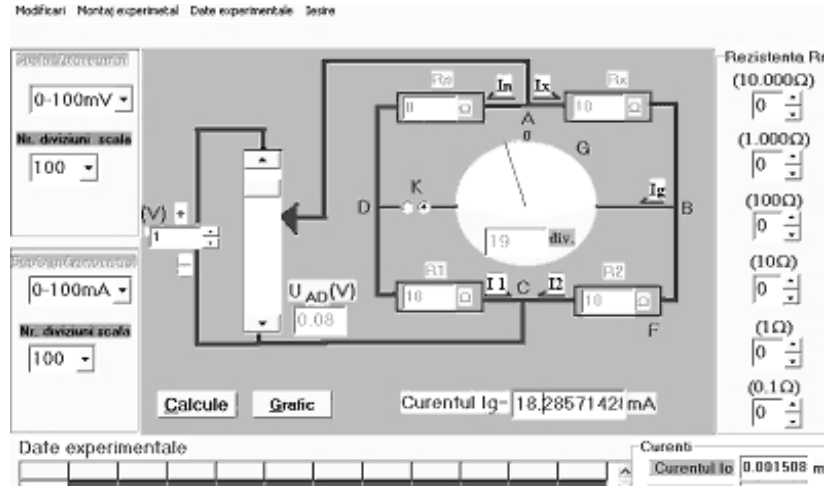

figure 6 - The computer interface for Wheatstone assembly

The electric appliance scheme is presented in Figure 5. After some needed initializing operations executed as in the real experiment, the students have to realize a balancing for the Whetstone's bridge modifying the $R_n$ resistance value.

The computer interface image for this program is presented in Figure 6. This application's part was designed to be very schematic, and, in the same time, a suggestive and faithful representation. The application initialization procedure requires same steps as in the real experiment.

The next example is of third generation software and allows a comparative voltmeter study. This computer appliance is a good circumstance to individual study of voltmeters' proprieties. The proper impedance, the sensibility and the value domains can be modified. One considered aspect is to obtain the device static function characteristics of the

$$\alpha = \alpha(U) \qquad (2)$$

where is de device indication.

The considered electrical device scheme is presented in Figure 7 and the interface application is depicted in Figure 8.

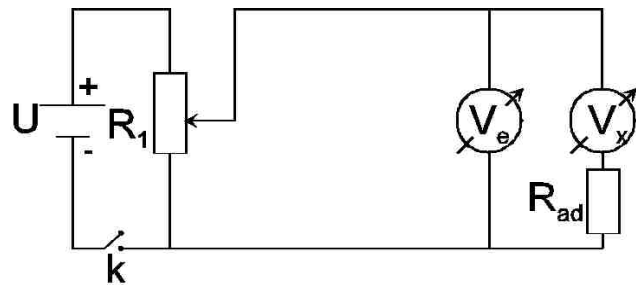

Figure 7 - The electric appliance for comparative voltmeter study

One highlighted aspect is the voltmeter sensibility

$$S = \frac{\Delta\alpha}{\Delta U} \qquad (3)$$

where represents the device indication change and the correspondent voltage variation. Students should estimate the sensibility value for different cases.



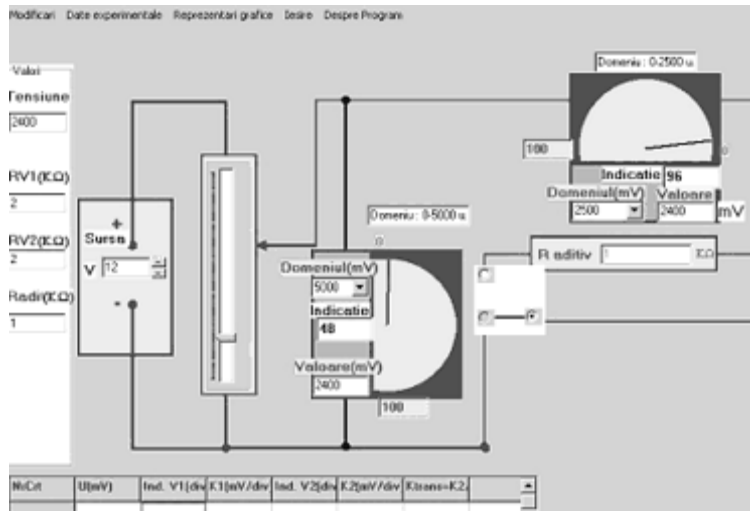

Figure 8 - The computer interface for comparative voltmeter study

For other applications, it could conjure up some more examples like the current measurement device study, or an experimental measure with an opposition method for the electromotor tension source parameters Figure 9 and Figure 10 or a thread bridge study Figure 11.

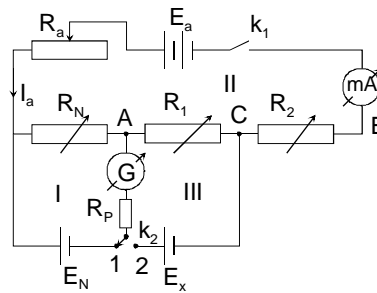

Figure 9 - The electric appliance for measure with an opposition method

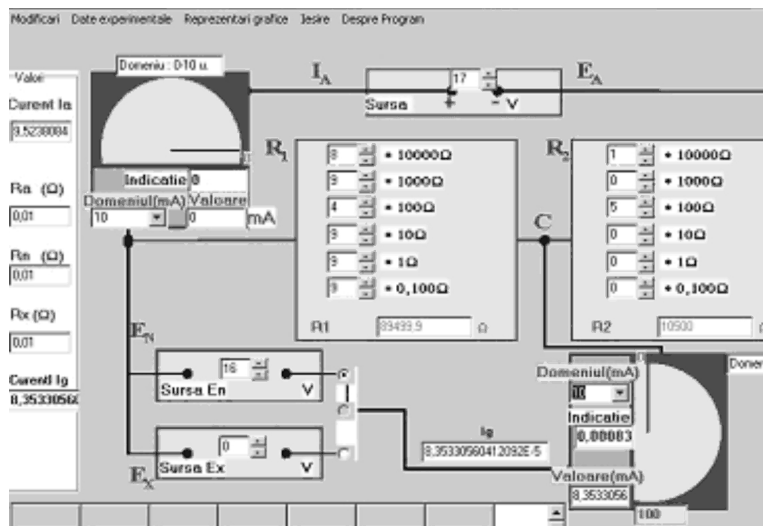

Figure 10 - The computer interface for measure with an opposition method



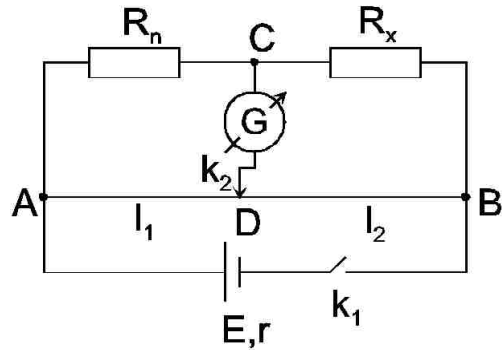

Figure 11 - The electric appliance for thread bridge study

In future papers will be presented other developed software applications for other experimental domains, as for the alternative current circuits for example or optics phenomena. The development requested more effort and more complex structures object.

All these programs asked a huge volume of work and of course, a long time for checking and testing the results. From this reason, a useful software application is hardly to be obtained faster.

## 4. Using Methods

First idea was to run the appropriate applications for every experimental work separately, in a demonstrative way, before the experimentation. During this operation, the teacher can focus on the specific aspects of the trial. He should mention some requested theory aspects and has to expose the needed operation for correct devices using.

All the programs are made respecting an important request: to may be tuned in order to simulate the specific parameters of present laboratory apparatus. In this way, every application can be set to be a faithful copy of the real devices. So the software can be run during the experiment for a comparative study.

A different way is to use them as a checking tool of the obtained results.

## Acknowledgments

The author wishes to thanks to A. M. Dariescu and C. Dariescu for help and fruitful discussions. Special thanks go to G. Puşcaşu for his advices to improve the applications. All the opinions in order to perfect these results are welcome and impatiently expected.